# High Sensitive Measurement of Absorption Coefficient and Optical Nonlinearities with a Single Experimental Setup

S. Abbas Hosseini, A. Sharan, D. Goswami[*]

Tata Institute of Fundamental Research, Homi Bhabha Road, Mumbai 400 005, India.

Accurate knowledge of absorption coefficient of a sample is a prerequisite for measuring the third order optical nonlinearity of materials, which could become a serious limitation for unknown samples. We introduce a new method, which measures both the absorption coefficient and the third order optical nonlinearity of materials with high sensitivity in a single experimental setup. We use a dual-beam pump-probe experiment under different conditions to achieve this goal. We also demonstrate a counterintuitive coupling of the non-interacting probe-beam with the pump-beam in pump-probe z-scan experiment.

---

[*] debu@tifr.res.in

There is a continued effort in making sensitive measurements on absorption coefficient ($a_0$) and the third-order optical nonlinearity ($c^{(3)}$). Most of the experimental techniques are focused on measuring one or the other of these two important parameters. A single experimental technique to measure both the parameters, however, is yet to emerge, which could be of significance in the study of new materials. In this letter we introduce a single experimental technique to measure both the parameters. We use communication relevant laser wavelength of 1560 nm for the work presented here.

One of the most important techniques to measure the real and imaginary parts of $c^{(3)}$ was shown by Sheik Bahaei et al.[1] This technique is simple and versatile and can measure the above parameters with high sensitivity. However, an accurate knowledge of $a_0$ is necessary for the use of this technique, which is a serious limitation for unknown samples. An easy way to measure $a_0$ is to use the Beer's law[2-3], which operates in the linear absorption regime and has limited sensitivity. More sensitive methods have been developed, of which the technique[4-10] using thermal lens (TL) effect is perhaps the simplest and the most effective. In this method, a lens focuses the laser beam into the sample resulting in a temperature gradient, which in turn produces a spatial gradient in refractive index. The relative change in transmittance of the laser beam can then be measured, after passing through an aperture, with the help of a detector[4-6]. Shen et al.[7-9] introduced a pump probe laser scheme under mode-mismatched and mode-matched conditions to improve sensitivity of the TL method. More recently, Marcano et. al.[10] have used this method to measure the absorption coefficient of water with high accuracy. Our aim has been to measure the absorption coefficient as well as the real and imaginary parts of $c^{(3)}$ with high sensitivity in a transparent sample using a single experimental setup.

Our experimental scheme involves a sub-100 femtosecond mode-locked Er:doped fiber laser (IMRA Inc.) operating at a repetition rate of 50 MHz and provides the fundamental (1560nm) and its second-harmonic wavelength (780nm) simultaneously as a single output. The sample is double distilled water in a 1.6 mm thick BK7-glass cuvette. To measure the real and imaginary parts of the third order nonlinear susceptibility in water, we filtered one of the two wavelengths and used each of them independently in the z-scan geometry. We used a 75 mm focusing lens, which results in the maximum intensity of 0.56 MW/cm$^2$ and 2.8 MW/cm$^2$ for the 1560nm and 780nm wavelengths respectively, at the focal point. We used a silicon photodetector (Thorlab: DET 210) for the 780nm beam detection, and an InGaAs photodetector (Acton Research) for the 1560nm beam detection.

We find that while the 1560nm beam produces the familiar z-scan results (fig.1), the 780nm beam is unable to produce any effect at our peak powers. This enabled us to use the 780nm wavelength as the non-interacting probe beam for the subsequent experiments where we use both the wavelengths from the laser simultaneously to measure $a_0$ (fig.2). This turns out to be the mode-mismatched pump-probe experiment since the 75mm lens focuses the 780nm probe beam to its minimal spot size position 0.4mm ahead of the pump beam of 1560nm. Furthermore, the focal spot size of 9ìm for 780nm is 15ìm smaller than the corresponding 1560nm spot size at the focus. However, our extremely sensitive experimental measurement schemes will also show that we are able to detect the minute effects of the pump beam on the probe beam when they co-propagate in the medium (fig.2).

For nonlinear materials the index of refraction $n$ is expressed in terms of nonlinear $n_2$ or $\gamma$ through the relation[1]: $n = n_0 + \frac{n_2}{2}|E|^2 = n_0 + \gamma I$, where $n_0$ is the linear index of refraction, $E$ is the peak electric field (cgs) and $I$ (MKS) is the intensity of the laser beam inside the sample. $n_2$ and $\gamma$ are related to each other as: $n_2(esu) = \frac{cn_0}{40\pi}\gamma(m^2/W)$, where $c$(m/s) is the speed of light in vacuum. The third order optical susceptibility is considered to be a complex quantity: $\chi^{(3)} = \chi_R^{(3)} + \chi_I^{(3)}$. The real and imaginary parts are related to the $\gamma$ and $\beta$ respectively[11] where $\beta$ is the nonlinear absorption coefficient and is defined as $\alpha(I) = \alpha_0 + \beta I$. The typical z-scan data with fully open aperture is insensitive to nonlinear refraction; therefore, the data is expected to be symmetric with respect to focus. For materials with multiphoton absorption, there is a minimum transmittance in focus (valley) and for saturable absorber samples, there is maximum transmittance in the focus (peak)[1]. Our sample shows a peak in the open aperture data (fig.1a) due to saturation in absorption at 1560nm. Saturation leads to thermal lensing as is seen in the pump-probe data of fig. 2b with 40% closed aperture collection of 780nm probe beam. We use saturation absorption relations to calculate the unknown parameters. The asymmetry in the experimental graphs with respect to the focal point is due to thermal effects generated by the high repetition rate (50MHz) of our laser in contrast to the two-photon absorption cases explored earlier[12].

Shen et. al.[7-9] have derived an expression for the TL signal using diffraction approximation for Gaussian beams in steady state case as:

$$S(z) = [1 - \frac{q}{2}\tan^{-1}(\frac{2mV}{1+2m+V^2})]^2 - 1 \quad (1)$$

where

$$m = (w_p/w_o)^2,$$

$$V = (z-a_p)/z_p + [(z-a_p)^2 + z_p^2]/[z_p(L-z)],$$

$$w_{p,o} = b_{p,o}[1+(z-a_{p,o})^2/z^2_{p,o}]^{1/2}, \quad (2)$$

$$q = -P_o a_0 l(ds/dT)/k l_p,$$

$z$ is sample position with respect to the focal point, $a_p$, $a_o$, $z_p$, $z_o$ and $b_p$, $b_o$, are position of the waists, the confocal parameters and the beam radius for the probe and pump beams, respectively. $l_p$ is the wavelength of the probe beam, $ê$ is thermal conductivity coefficient of the sample. $L$ is the detector position, $P_o$ is the total power of the pump beam and $l$ is the sample thickness. The value of $á_0$ is obtained from the closed aperture data. The solid line in the fig.2b is the result of a theoretical fit to Eq.1. This fit gives the value of phase shift, $è = 9.957$, which when substituted in Eq.2 with the parameters[10] $ds/dT = -9.1 \times 10^{-5} K^{-1}$ and $k = 0.598 \times 10^{-2} WK/cm$ for pure water, we get the calculated value for $a_0$ as 10.6327cm$^{-1}$ for the 1560nm beam.

We did the scan in fully open aperture case with 1560nm wavelength. The result is shown in fig. 1a. The solid line is the theoretical fit, which is derived by solving the differential equation for the transmitted light through a thin sample of thickness $l$ [13]

$$T(z) = \boldsymbol{h} + \frac{\boldsymbol{b} I_0 l}{(1 + z^2/z^2_0)} \qquad (3)$$

where $z_0 = k w_0^2/2$, $k$ and $w_0$ are the wave vector and the minimum spot size in the focal point respectively, while $\boldsymbol{h}$ and $\boldsymbol{b}$ are the fitting parameters. The best fit gives the value of $\boldsymbol{b} = -2.58$ cm/GW.

Thus $\boldsymbol{a}_0$ and $\boldsymbol{b}$ value of the sample at 1560nm wavelength are known. Next we collected z-scan data by closing 40% of the aperture. The result is shown in fig. 1b. The valley-peak structure suggests a self-focusing effect inside the sample. By replacing the values of $\boldsymbol{a}_0$ and $\boldsymbol{b}$ in the theoretical model[1] which is derived by using ray optics distortion inside a nonlinear thin sample we calculate $\gamma = 1.57 \times 10^{-3}$ cm$^2$/GW, which is proportional to $n_2 = 4.9 \times 10^{-12}$ esu.

As mentioned earlier, when only the 780nm beam passes through the sample, there is a constant signal and we do not see any peak or valley. The absorption coefficient of water at 780nm is so small that at our intensities, there is no saturation effect. However, when 1560nm beam is simultaneously present, it affects the propagation of the 780nm beam. As the 1560nm beam starts to saturate the sample at its focal point, the 780nm beam also experiences a saturated environment, whereby its transmittance increases at its focal point (fig.2a). Thus, this technique can be used to measure the nonlinear absorption coefficient of materials, whose linear absorption coefficients are extremely small. We fit the experimental data using a theoretical expression given by Eq. 3. The solid line in fig.2a is the theoretical fit. The calculated parameter $b$ = -8.5×10$^{-3}$ cm/GW is extremely small and indicates the high sensitivity of our approach.

We have been able to estimate both the linear absorption coefficient as well as the real and imaginary parts of third order optical susceptibility of water with the pump-probe technique and the z-scan technique in a single experimental setup. This helped us in retaining the sensitivity of the conventional z-scan measurement while enhancing the versatility of such measurements to unknown samples.

[1] M. Sheik-Bahae, A.A. Said, T. Wei, T.H. Hagand, D.J. Hagan, and E.W. Van Stryland, IEEE 760 (1990).

[2] G.T. Fraser, A.S. Pine, W.J. Lafferty and R.E. Miller, J. Chem. Phys. **87**, 1502 (1987).

[3] H. Petek, D.J. Nesbitt, D.C. Darwin and C.B. Moore, J. Chem. Phys. **86**, 1172 (1987).

[4] N.J. Dovichi and J.M. Harris, Anal. Chem. **51**, 728 (1979).

[5] G.R. Long and S.E. Bialkowski, Anal. Chem. **56**, 2806 (1984).

[6] N.J. Dovichi and J.M. Harris, Anal. Chem. **53**, 106 (1981).

[7] J. Shen, M.L. Baesso and R.D. Snook, J. Appl. Phys. **75**, 3738 (1994).

[8] J. Shen, R.D. Lowe and R.D. Snook, Chem. Phys. **165**, 385 (1992).

[9] J. Shen, A.J. Soroka and R.D. Snook, J. Appl. Phys. **78**, 700 (1995).

[10] A. Marcano O., C. Loper and N. Melikechi, Appl. Phys. Lett. **78**, 3415 (2001).

[11] P. Gunter, "Nonlinear Optical Effects and Materials", Springer, Berlin, (2000).

[12] T.D. Krauss and F.W. Wise, Appl. Phys. Lett. **65**, 1739 (1994).

[13] S. Vijayalakshmi, F. Shen and H. Gerbel, Appl. Phys. Lett. **71**, 3332 (1997).

**Figure captions:**

**Fig. 1.** Measured z-scan of a 16mm thick double distilled water using 95 femtosecond pulses at λ=1560nm (diamond) and theoretical fit (solid line) for **(a)** fully open aperture, and **(b)** 40% closed aperture.

**Fig. 2.** Measured z-scan transmittance of 80fs pulses of λ=780nm as a probe through a 16mm thick double distilled water being irradiated with 95fs of λ=1560nm as pump (diamond) and theoretical fit (solid line) in **(a)** fully open aperture and **(b)** 40% closed aperture.

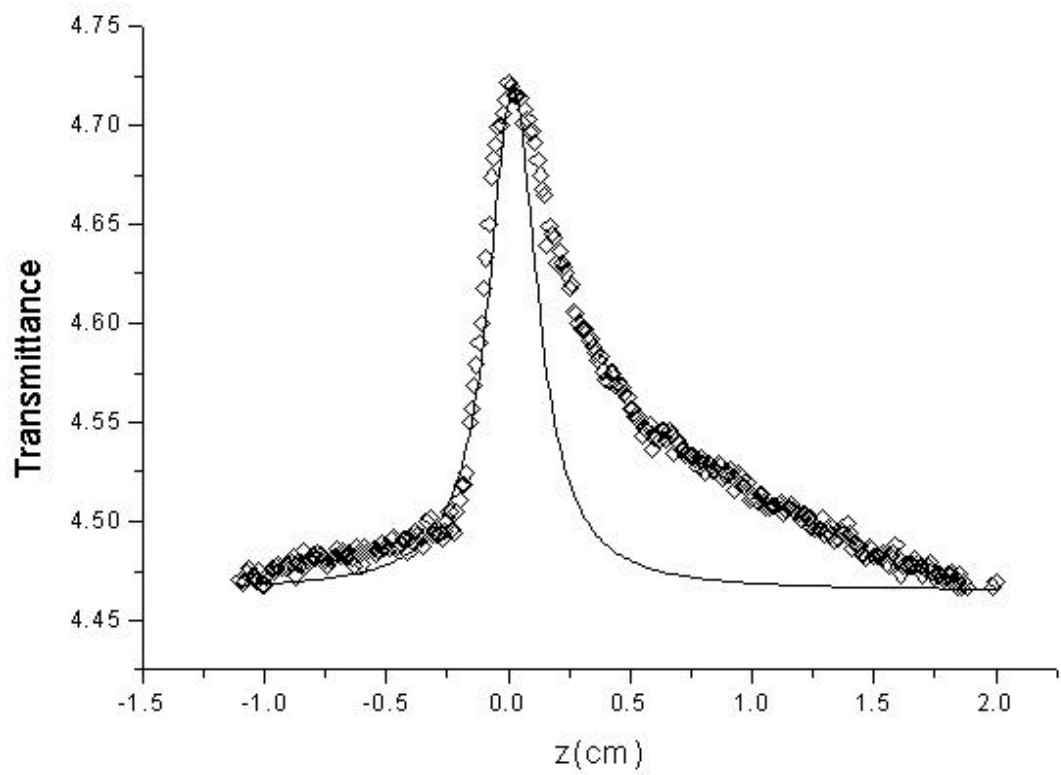

**Fig. 1a**

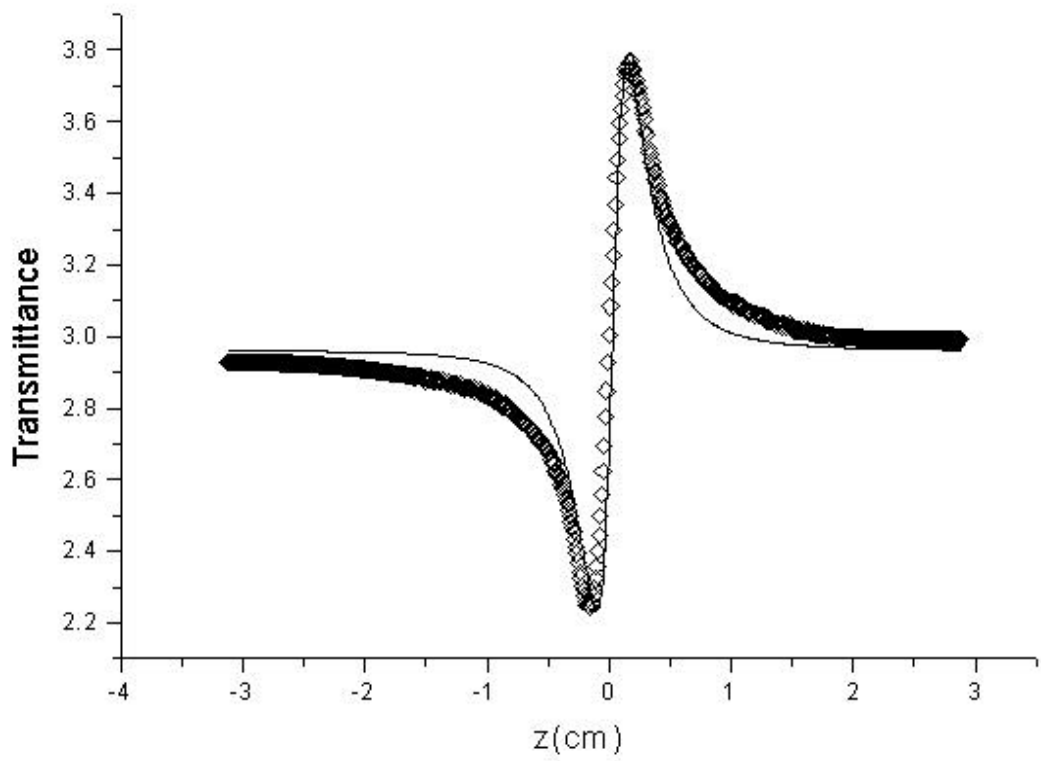

**Fig. 1b**

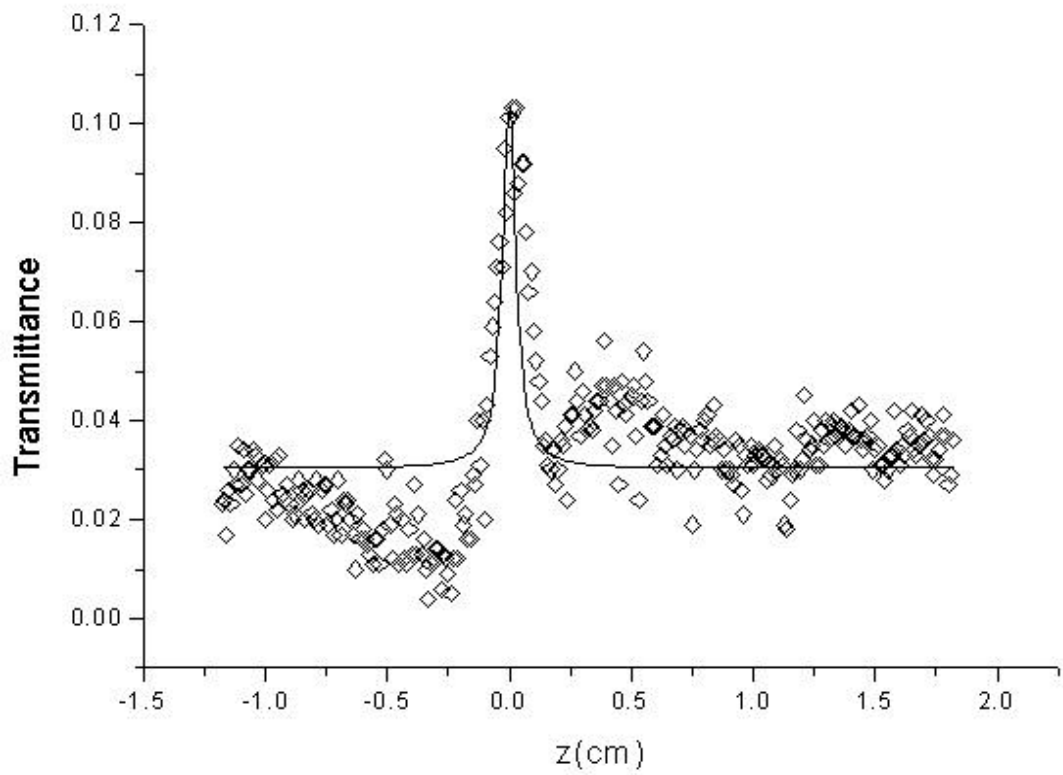

**Fig. 2a**

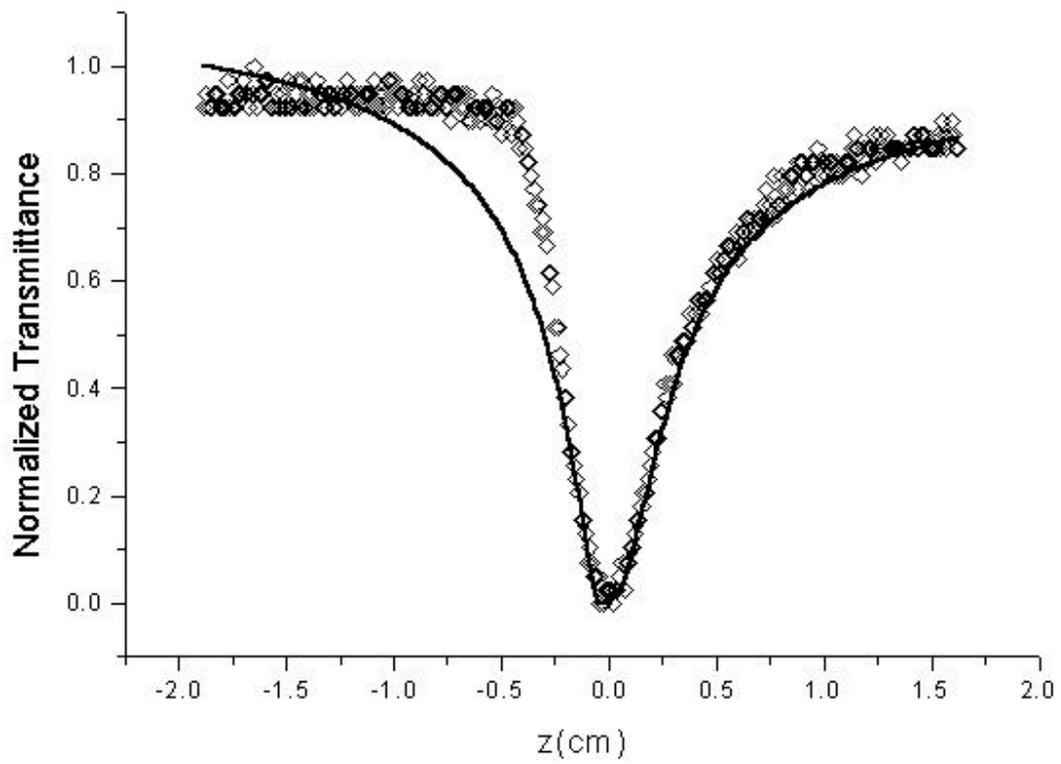

**Fig. 2b**